\documentclass[journal=amlcef,manuscript=letter]{achemso}
\setkeys{acs}{articletitle=true}

\usepackage{graphicx}
\usepackage{amsmath}
\usepackage{amssymb}
\usepackage{booktabs}
\usepackage{xcolor}
\usepackage{url}

\newcommand{\aff}{AFF-r2SCAN}
\newcommand{\affs}{AFF-r2SCAN-smooth}
\newcommand{\gbh}{GB10-hours}

\title{Scaling a Universal Graph Neural Network Force Field to Billions of Atoms on Desk-Side GPUs}

\author{Jaehyung Lee}
\affiliation{Department of Materials Science and Engineering, Johns Hopkins University, Baltimore, Maryland 21218, United States}
\author{Kamal Choudhary}
\email{kchoudh2@jh.edu}
\affiliation{Department of Materials Science and Engineering, Johns Hopkins University, Baltimore, Maryland 21218, United States}
\alsoaffiliation{Department of Electrical and Computer Engineering, Johns Hopkins University, Baltimore, Maryland 21218, United States}

\begin{document}

\begin{tocentry}
\includegraphics[width=3.25in]{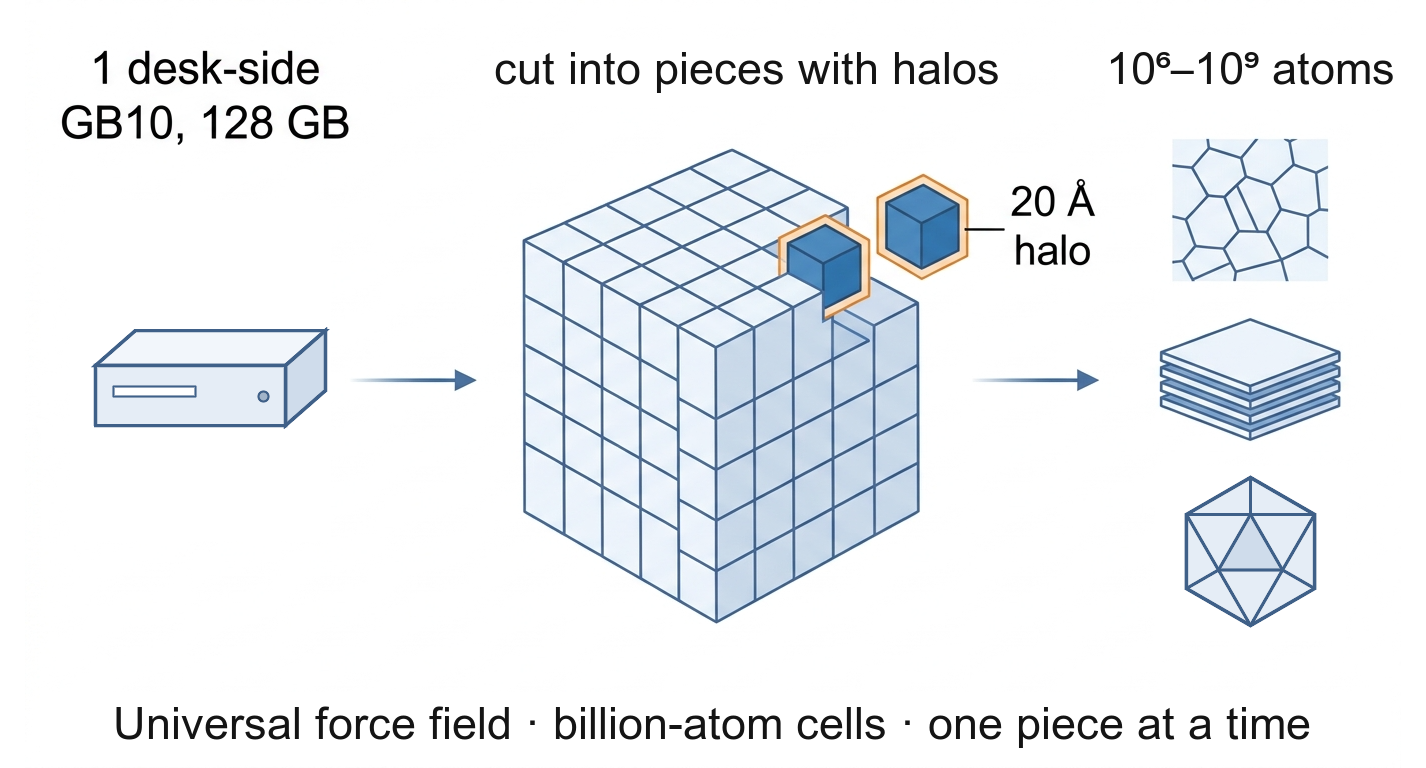}
\end{tocentry}

\begin{abstract}
Machine-learning force fields are capable of simulating billions of atoms, but usually require high-performance supercomputing clusters. We show that a universal graph neural network force field, ALIGNN-FF, evaluates energies and forces of more than one billion atoms on desk-side computers such as the NVIDIA GB10, which has 128~GB of unified memory, by using a cube-and-halo decomposition. The halo contains every atom within the model's 20~\AA{} receptive field, which makes the result exact. One GB10 holds 216,000 atoms with forces as a single graph. The cube-and-halo decomposition removes this limit, enabling an evaluation of 2 billion atoms on nine GB10s. We also demonstrate three case studies: a billion-atom polycrystal, a 102-million-atom microelectronic gate stack and a 1.07-million-atom satellite tobacco mosaic virus in water. This brings single-point evaluations at experimentally relevant sizes within reach of one research group. The code is available at: \url{https://github.com/atomgptlab/scalingmat}.
\end{abstract}

Many properties of materials are determined by mesoscale features that span nanometers to micrometers: the grain boundaries of a polycrystal, the interfaces of a transistor, the packing of proteins and lipids in a virus~\cite{li2018mesoscience,natmater2012interface,suo2026trillion}. Simulating such structures atom by atom takes $10^{8}$ to $10^{12}$ atoms, far beyond density functional theory (DFT). Machine-learning force fields (MLFFs) trained on DFT data close most of this gap~\cite{behler2007nnpotential,bartok2010gap,thompson2015snap,unke2021mlff,choudhary2022dlreview}, because they retain much of the accuracy of their DFT training data at a cost that grows linearly with the number of atoms. The largest MLFF calculations show how far this has gone: 127 million and then billions of Cu atoms with Deep Potentials on Summit~\cite{jia2020deepmd,guo2022deepmd}, one billion carbon atoms with SNAP on 27,900 GPUs~\cite{nguyencong2021snap}, a solvated HIV capsid of 44 million atoms with Allegro~\cite{musaelian2023allegro,kozinsky2023allegro}, 29 billion water atoms on the new Sunway machine~\cite{wang2025sunway} and 1.62 trillion Cu atoms with a neuroevolution potential (NEP) on 45,000 accelerators~\cite{fan2021nep,suo2026trillion}. These records share two features. Each ran on a leadership-class machine, and each used a potential fitted to one system or one class of systems.

\begin{figure}[htbp]
\includegraphics[width=\textwidth]{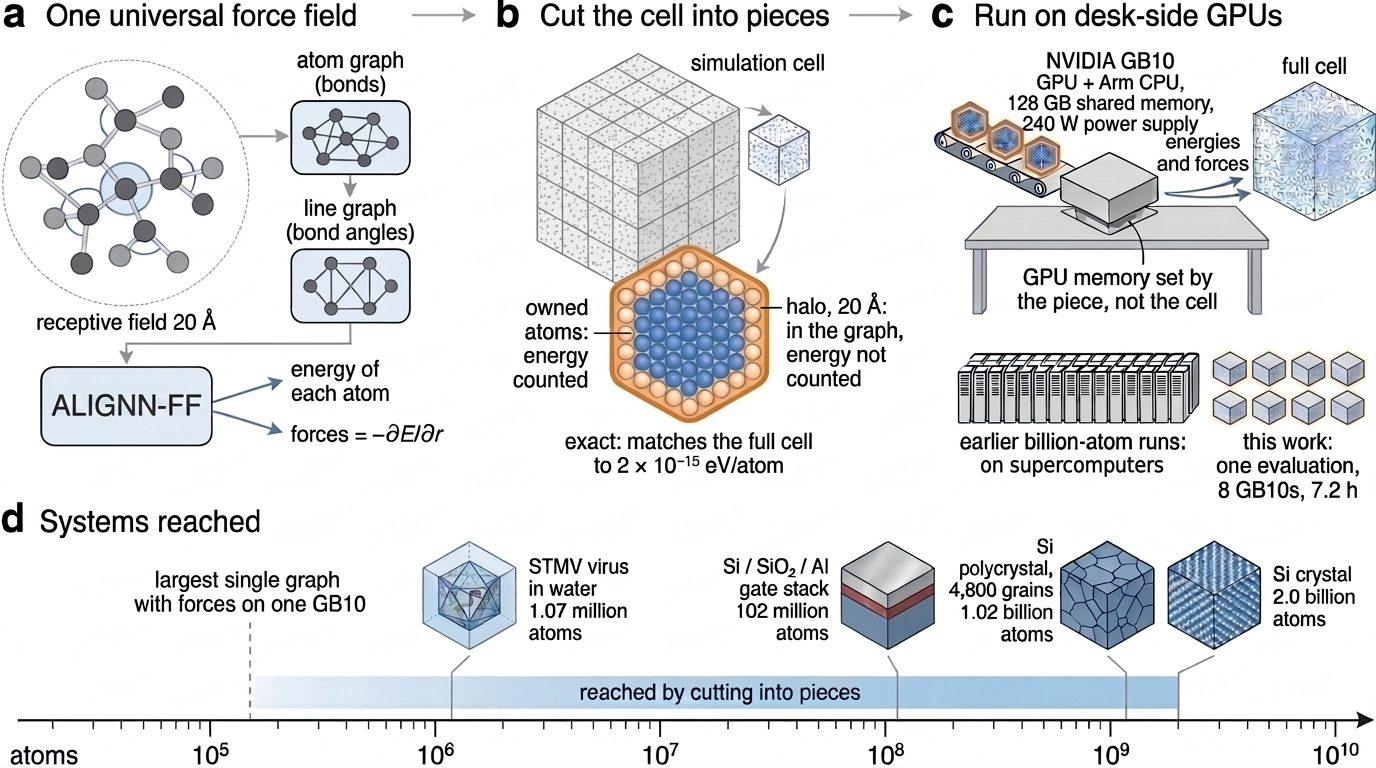}
\caption{Overview. (a) ALIGNN-FF represents a structure as a graph of atoms joined by bonds and a line graph of bond angles, and predicts the energy of each atom; forces are the negative gradient of the total energy. An atom's energy depends only on atoms within its receptive field, 20~\AA{} for \affs{}. (b) The cell is cut into cubes, and each cube is evaluated with a halo of every atom within 20~\AA{} of it. Only the cube's own atoms contribute energy; the agreement with the full cell quoted in the panel is in double precision, and in single precision the difference is float32 roundoff. (c) The pieces run one at a time on a desk-side GB10, and their energies and forces are summed into the full cell, so GPU memory is set by the size of a piece rather than of the cell. (d) Structures evaluated with energies and forces in this work, placed approximately by atom count, against the largest single graph with forces on one GB10 (216,000 atoms). The run of the satellite tobacco mosaic virus (STMV) measures cost only. The structures are drawn schematically.}
\label{fig:overview}
\end{figure}

Universal force fields remove the second requirement. ALIGNN-FF~\cite{choudhary2021alignn,choudhary2023alignnff}, M3GNet~\cite{chen2022m3gnet}, CHGNet~\cite{deng2023chgnet} and MACE-MP-0~\cite{batatia2023macemp0} are trained once on DFT data spanning the periodic table~\cite{choudhary2020jarvis,garrity2021wannier,choudhary2025jarvisinfra} and apply to new chemistries without refitting. They are graph neural networks with several message-passing layers and, in ALIGNN-FF's case, explicit three-body (bond-angle) messages, so they cost far more per atom than SNAP or NEP and are usually run on $10^{2}$ to $10^{5}$ atoms. The NVIDIA GB10 Grace-Blackwell superchip~\cite{nvidia2025dgxspark} changes the hardware side. It pairs a Blackwell GPU with a 20-core Arm CPU and 128~GB of memory that both share, in a desk-side system with a 240~W power supply. Its memory per device is close to that of data-center GPUs, at a fraction of their size, power and price, though with an order of magnitude less memory bandwidth (273~GB/s). This raises a simple question: how many atoms can a universal force field handle on hardware a single research group can own? Here we answer it for ALIGNN-FF (Figure~\ref{fig:overview}). The decomposition we use is established; what is new is the system size it reaches with a universal model on such hardware.

ALIGNN-FF represents a structure as a graph of atoms joined by bonds within 5~\AA{} (at most 52 neighbors per atom), plus a line graph whose nodes are bonds shorter than 3.5~\AA{} and whose edges are bond angles. These three settings shape every result below. The short three-body cutoff keeps the line graph small, 16 line-graph edges per atom in Si (pairs of bonds that share an atom, including each bond paired with its reverse), where every bond within 5~\AA{} would give 784; M3GNet and CHGNet restrict their three-body terms in the same way, to 4 and 3~\AA{}~\cite{chen2022m3gnet,deng2023chgnet}. The two cutoffs and the number of layers fix the distance over which one atom can affect another, the receptive field, which is 20~\AA{} for the published \aff{} model~\cite{choudhary2026matpesr2scanff}, trained on MatPES r2SCAN data~\cite{kaplan2025matpes}. The forces of \aff{} jump where a bond crosses the 3.5~\AA{} three-body cutoff, which we found and removed as described below, so every large-scale result in this work uses \affs{}: the same network with smooth cutoffs, trained from scratch on the same data. We first measured the largest pristine Si cell that one GB10 evaluates as a single graph. Memory per atom is constant, so the limit is set by memory alone: 551,368 atoms for the energy (0.195~MB/atom) and, with forces, 54,872 atoms through the ASE calculator, 140,608 through a single \texttt{forward\_\allowbreak tensors} call and 216,000 with layer checkpointing (0.550~MB/atom, 10.2~s; 119~GB of the 130.7~GB available, so the next size, 238,328 atoms, would not fit), which recomputes each layer's activations during the backward pass instead of storing them. Three larger ALIGNN-FF models, with four ALIGNN and four graph-convolution layers and twice the hidden width, stop at 238,328 to 262,144 atoms without forces and 97,336 with forces (Table~S2).

\begin{figure}[htbp]
\centering
\includegraphics[width=0.5\textwidth]{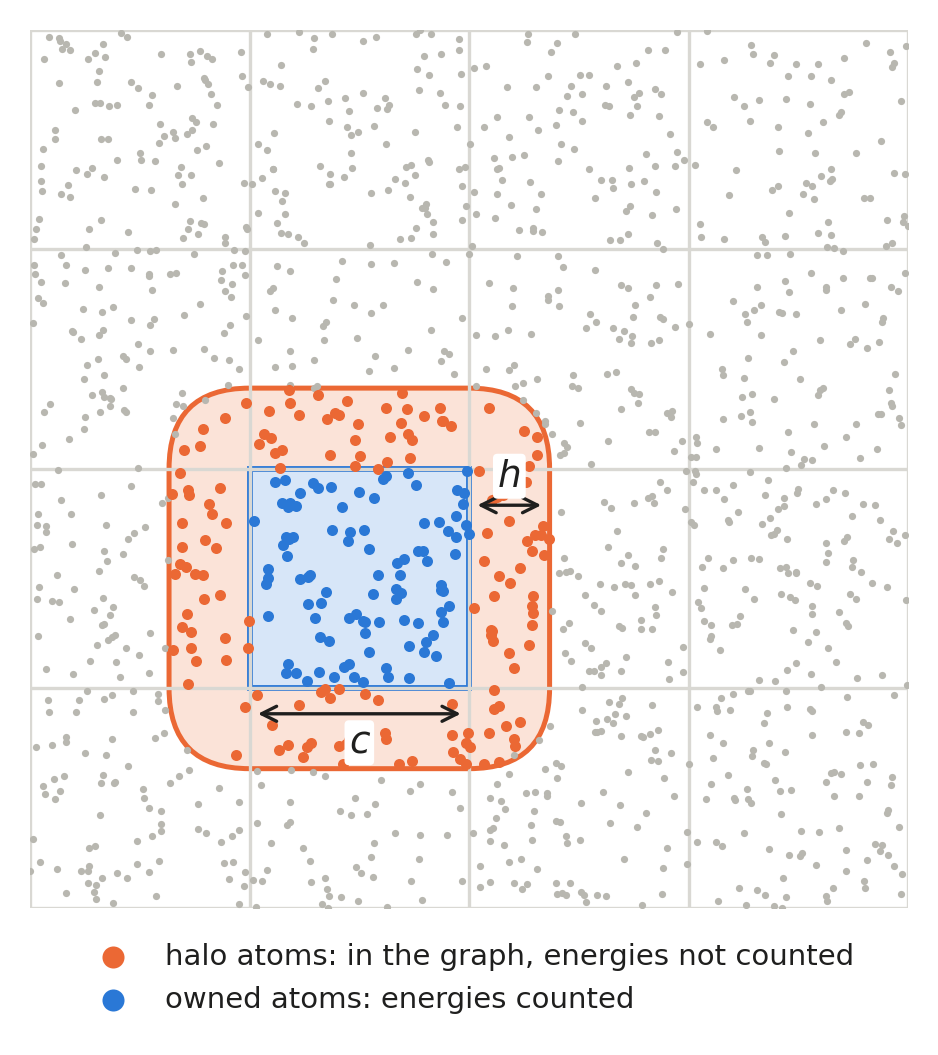}
\caption{Cube-and-halo decomposition, in two dimensions. The cell is cut into cubes of side $c$. Each piece is one cube (blue) plus every atom within $h$ of it (orange), evaluated as an isolated cluster. Only the cube's own atoms contribute energy; the forces from that energy are added to every atom in the piece. With $h$ at least the model's receptive field, and the 52-neighbor cap not reached, the sum over pieces equals the full-cell energy and forces.}
\label{fig:decomp}
\end{figure}

Because an atom's energy depends only on atoms within the receptive field, a cell can be cut into $k^{3}$ cubes and each cube evaluated together with a halo of every atom within distance $h$ of it (Figures~\ref{fig:overview}b and \ref{fig:decomp}). This is the spatial decomposition with ghost atoms of parallel MD codes~\cite{thompson2022lammps}, with the ghost region widened from one cutoff to the receptive field, as message-passing force fields require~\cite{park2024sevennet,han2025distmlip}. Only the cube's own atoms count toward the energy, and the gradient of that partial energy with respect to every atom in the piece, halo atoms included, is added to a full-cell force array. If $h$ is at least the receptive field and no atom has more neighbors within the cutoff than the 52-neighbor cap, the sum over pieces is the full-cell result, and the only difference is the order of floating-point additions. On 8,000-atom cells that fit as a whole, chunked and full-cell results agree to $2\times10^{-15}$~eV/atom and $4\times10^{-13}$~eV/\AA{} in double precision. In single precision, used for all production runs, they differ by up to $9\times10^{-7}$~eV/atom and $3\times10^{-4}$~eV/\AA{}, which is float32 roundoff (Figure~S2). The pieces are independent, so they run one after another on one GB10 or on any number of them, with a single sum at the end and no communication in between. The price is that halo atoms are evaluated more than once. Schemes that exchange atom and bond features between devices during the forward pass avoid that recomputation~\cite{park2024sevennet,han2025distmlip}, but the devices must then communicate at every layer.

\begin{table}[htbp]
\caption{Chunked evaluations with \affs{} on GB10s, in single precision. ``GB10s (tasks)'' is the number of machines that ran the tasks; ``Largest piece'' is the atom count (cube plus halo) of the largest piece and ``GPU'' its peak GPU memory. ``Total'' is the summed wall time of all tasks, which is the time on one GB10. Si cells have every atom displaced by 0.05~\AA{} per component; the Si cells of $10^{9}$ atoms and more were generated from the lattice piece by piece. None of the runs reached a memory limit, so these are the largest sizes tried, not maxima. STMV is the satellite tobacco mosaic virus. The same runs with \aff{} are given in Table~S3 and the text after it.}
\label{tab:runs}
\small
\setlength{\tabcolsep}{4pt}
\begin{tabular}{lrcrrrr}
\toprule
System & Atoms & Forces & GB10s & Largest & GPU & Total \\
 & & & (tasks) & piece & (GB) & (\gbh) \\
\midrule
Si & 8,000,000 & yes & 1 (1) & 106,067 & 47.5 & 0.29 \\
Si & 64,000,000 & yes & 6 (6) & 106,054 & 47.5 & 2.34 \\
Si & 1,000,000,000 & no & 8 (10) & 261,222 & 49.3 & 9.29 \\
Si & 1,000,000,000 & yes & 8 (12) & 71,578 & 31.9 & 43.2 \\
Si & 2,000,376,000 & no & 9 (10) & 328,336 & 62.1 & 17.7 \\
Si & 2,000,376,000 & yes & 9 (20) & 104,194 & 46.6 & 72.3 \\
Si polycrystal, 4,800 grains & 1,021,299,267 & yes & 9 (10) & 69,056 & 30.8 & 45.1 \\
Si/SiO$_2$/Al gate stack & 102,361,540 & yes & 3 (6) & 39,059 & 52.8 & 10.7 \\
STMV virus in water & 1,066,628 & yes & 1 (1) & 22,263 & 58.3 & 0.92 \\
\bottomrule
\end{tabular}
\end{table}

\begin{figure}[htbp]
\includegraphics[width=\textwidth]{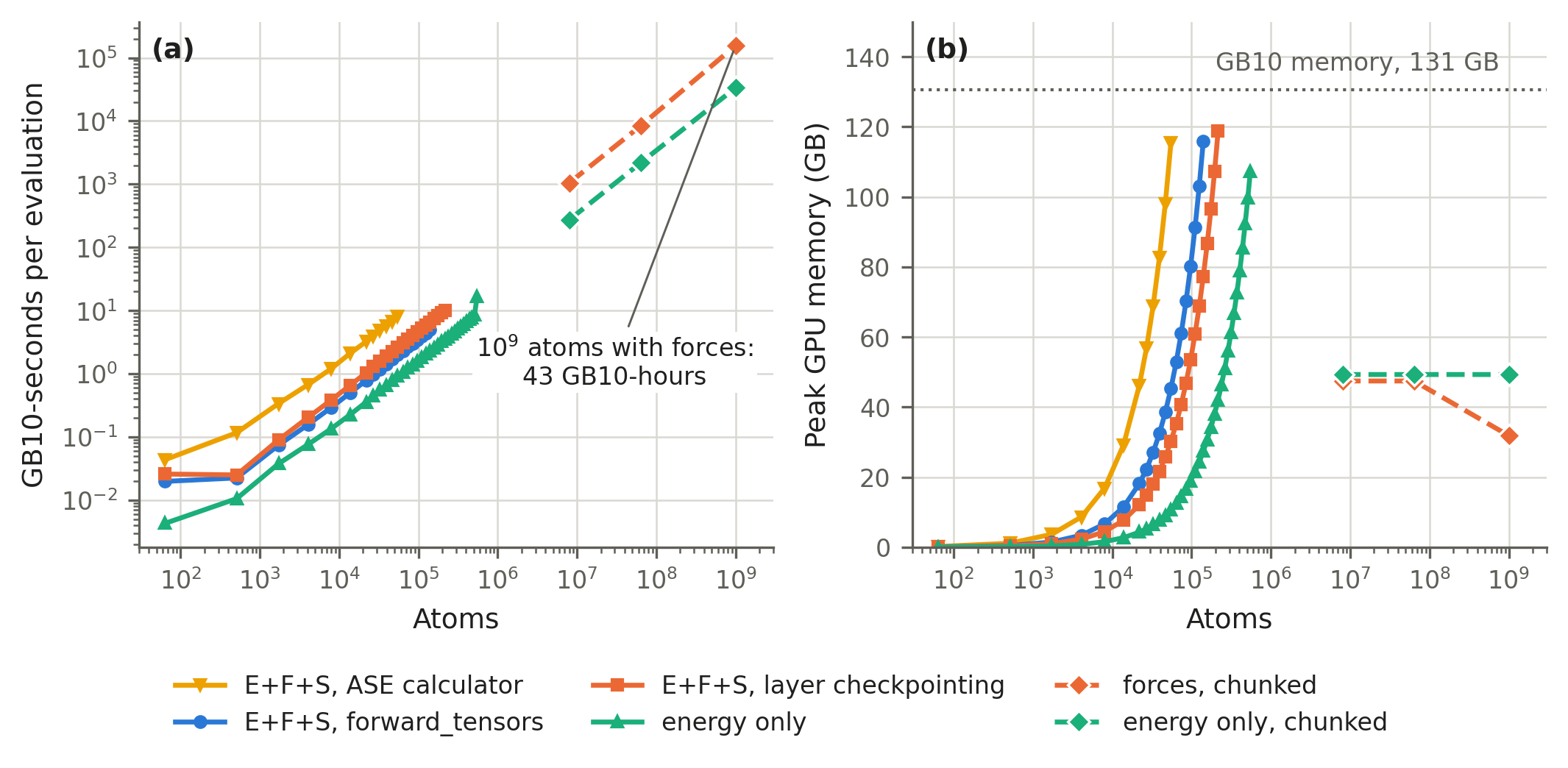}
\caption{Cost and memory of \affs{} on GB10, from 64 atoms to one billion. (a) GB10-seconds per evaluation: median time of one call for single graphs (solid), summed time of all tasks for chunked runs (dashed). The last energy-only point is slowed by graph construction near the memory limit. (b) Peak GPU memory. Single graphs grow linearly in memory until the host stops the process; chunked runs stay at the memory of one piece, set by the chosen piece size. The billion-atom force run used smaller pieces than the others, hence its lower memory. Pristine Si for single graphs, rattled Si for chunked runs. The dotted line in (b) is the memory available to PyTorch, 121.7~GiB (130.7~GB), of the 128~GiB the CPU and GPU share.}
\label{fig:scaling}
\end{figure}

With the decomposition, a larger cell is cut into more pieces of the same size, so it takes more time but no more GPU memory (Table~\ref{tab:runs}, Figure~\ref{fig:scaling}). One GB10 evaluated eight million Si atoms in 4.5 minutes for the energy and 17.5 minutes with forces. At 64 million atoms each force task did 1.33 times the work of the eight-million-atom run and took 1.32 to 1.39 times as long, so the cost is linear in the number of atoms. At one billion atoms, a Si cube 0.27~$\mu$m on a side, the energy took 9.3 \gbh{} in ten tasks and the energy and forces 43.2 in twelve. With forces the slowest task took 3.7~hours, and no task needed more than 31.9~GB of GPU memory, for its largest piece of 71,578 atoms. Chunking costs throughput, because halo atoms are evaluated but not counted: one GB10 processes 21,100 atoms per second with forces on a single graph of $10^{5}$ atoms, 7,600 when chunking eight million atoms and 6,400 at one billion.

Past a billion atoms the limit is host memory, not GPU memory: a task that reads the cell from disk holds all of it, 62.1~GB for the billion-atom polycrystal below, in the 128~GB that the CPU and GPU share. For a crystal, each piece can instead be generated from the lattice when it is evaluated, with each atom's random displacement drawn from a hash of its index so that every piece displaces it the same way. On 8,000 atoms in double precision this reproduces a full-cell evaluation to $3\times10^{-13}$~eV/\AA{}. The billion-atom Si runs above were made this way, with at most 4.8~GB of host memory per task, and so was a 0.34~$\mu$m cube of 2,000,376,000 Si atoms, which took 17.7 \gbh{} for the energy and 72.3 with forces, with at most 5.7~GB. The two runs cut the cell into different pieces, and their energies agree to $3\times10^{-10}$~eV/atom.

The cost per atom depends on the structure. Diamond Si is a favorable case, with 16 bond angles per atom; FCC Al and Cu have 144 and liquid-density water 258. Chunking widens the gap, because a piece of a denser structure holds its atoms in a smaller volume while the halo stays 20~\AA{} wide: a water piece evaluates 12 atoms for every atom it owns, against 2.8 for Si. From samples of about a million atoms on one GB10, one force evaluation of a billion atoms would take 38 \gbh{} for Si (the billion-atom run took 43, with smaller pieces), 168 for Al, 255 for Cu and 643 for water (Table~S4). Throughput rises with piece size, so the largest piece that fits is the fastest (Table~S8). In water the neighbor cap binds, so a 20~\AA{} halo is no longer exact. The error it leaves is $4\times10^{-9}$~eV/\AA{}, about five orders of magnitude below single-precision roundoff, and a 25~\AA{} halo removes it.

\begin{figure}[htbp]
\includegraphics[width=\textwidth]{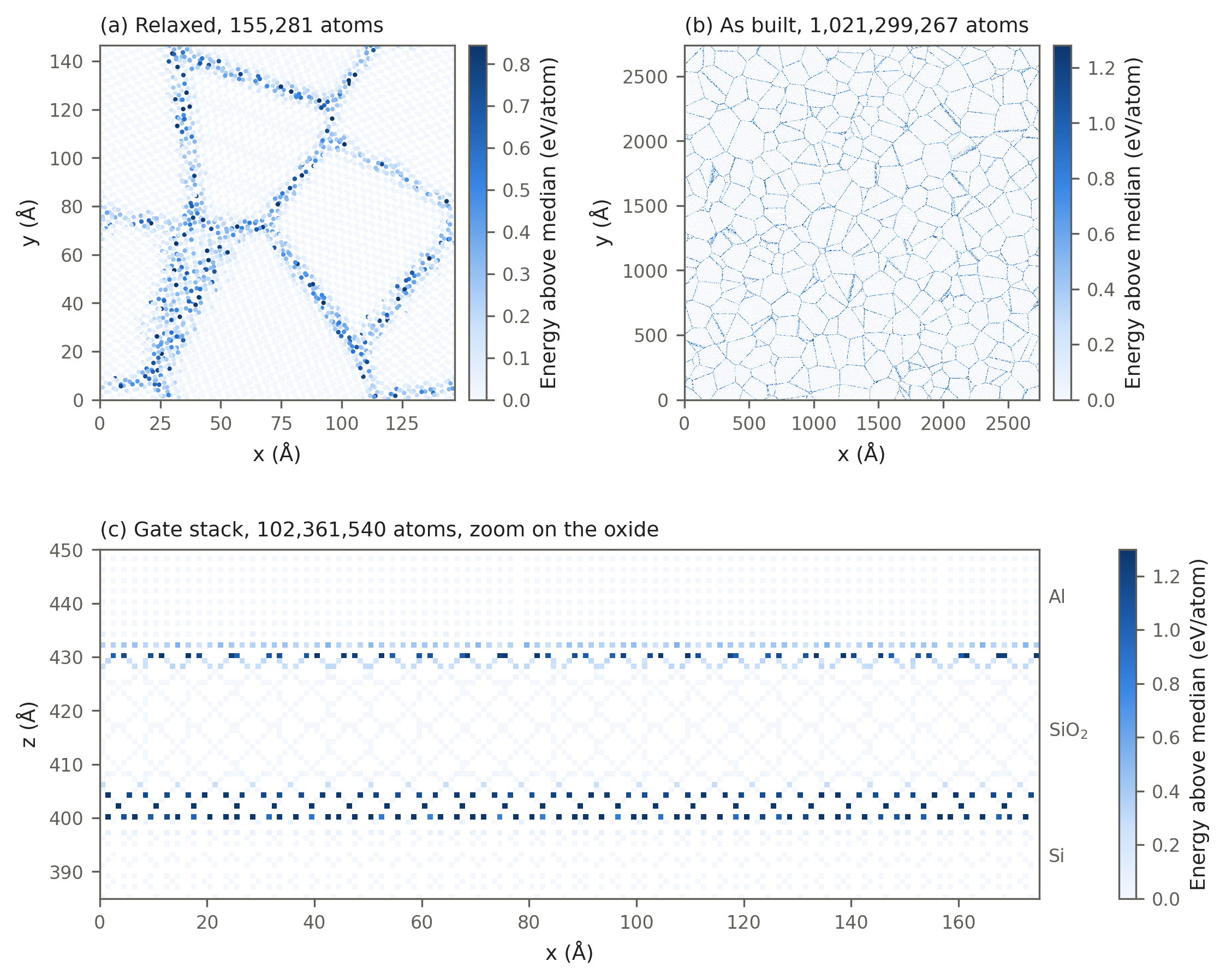}
\caption{Energy of every atom, relative to the median of its material, in 5~\AA{} slabs. (a) 8-grain Si polycrystal relaxed at fixed cell with FIRE until every force was below 0.05~eV/\AA{} (658 steps), atom by atom. (b) The billion-atom Si polycrystal as built, 2~\AA{} pixels. (c) Cross-section of the Si/SiO$_2$/Al gate stack around the oxide, 1~\AA{} pixels, with a separate median for each element in each layer. Grain boundaries and interfaces appear dark, and atoms below the median are drawn as zero. Per-atom energies are a partition defined by the model, not an observable, and are used here only to locate disorder.}
\label{fig:maps}
\end{figure}

We evaluated three structures as built, with energies and forces of every atom (Table~\ref{tab:runs}, Figure~\ref{fig:maps}). A Si polycrystal of 1,021,299,267 atoms, with 4,800 grains about 20~nm across in a 0.27~$\mu$m periodic cube, took 45.1 \gbh{} on nine GB10s. Its energy is 56.3~meV/atom above that of the perfect crystal, against 123~meV/atom for an as-built 8-grain cell whose grains are 2.2 times smaller, as expected if the excess follows the boundary area per atom. A planar gate stack of 102,361,540 atoms, 40~nm of Si, 2.9~nm of SiO$_2$ and 20~nm of Al in a 0.17~$\mu$m square cell, took 10.7 \gbh{} on three GB10s. Relative to the median of each element in each layer, the mean energy differs by more than 1~meV/atom only within 0.4 to 0.7~nm of the interfaces. The complete all-atom satellite tobacco mosaic virus (STMV) in water~\cite{freddolino2006stmv}, 1,066,628 atoms, took 0.92 \gbh{} on one GB10. This last run measures size and cost and gives no biological result, because the model was trained on inorganic materials and has not been validated on biomolecules.

With the published \aff{}, the 8-grain Si polycrystal of 155,281 atoms in Figure~\ref{fig:maps}a did not relax: FIRE~\cite{bitzek2006fire} at fixed cell, with the cell as one graph, ran 2,000 steps in 3.8~hours on one GB10. The energy fell by 49.8~meV/atom, 97\% of it within the first 100 steps, but the largest force stopped at 0.50~eV/\AA{}, ten times the target. The cause is the model, not the optimizer. The line graph keeps a bond only if it is at most 3.5~\AA{} long, with no smoothing of the angle terms, and moving a bond across that length by 0.0004~\AA{} changed the forces by 0.47 to 0.51~eV/\AA{}, against 0.005~eV/\AA{} for the same move on a bond far from any cutoff. Of the 20 atoms with the largest forces, 18 have a bond within 0.01~\AA{} of 3.5~\AA{}. Dynamics does not conserve energy either: in 2,000 constant-energy steps of 512 Si atoms started at 1500~K the total energy rose by 24.5~meV/atom, and a 1,536-atom water cell started at 300~K heated past 7,000~K within a picosecond.

We removed the jump by weighting each bond message and each angle message with a switch that falls smoothly to zero at its cutoff, and retraining with the settings and data of \aff{}, 100 epochs on one GB10 (Supporting Information). On the 992 MatPES frames in both test sets, which neither model saw in training, the mean absolute errors of this model, \affs{}, are 45.7~meV/atom and 0.144~eV/\AA{}, against 52.0 and 0.207 for \aff{} (Table~S6). With it the same relaxation converges, in 658 steps and 1.35~hours, with every force below 0.05~eV/\AA{} (Figure~\ref{fig:maps}a); the bond-crossing test gives the ratio of a smooth function at both cutoffs; and the energy drift in the two dynamics runs falls to 0.08 and 0.9~meV/atom. The three larger published ALIGNN-FF models, including the current default, show the same jump at 3.5~\AA{} (Table~S7). The switches cost time and memory: the runs of Table~\ref{tab:runs} take 7 to 8\% longer with forces than with \aff{} and about 4\% longer for the energy alone, and use 5 to 8\% more GPU memory with forces (Table~S3).

GPU memory is no longer the limit; time is. On one graph, memory sets a hard ceiling of about 216,000 atoms with forces, however long the calculation runs. With the decomposition the GPU holds one piece at a time, so a larger cell needs more pieces, not more memory, and what grows is the time: 0.29, 2.3 and 43 \gbh{} at eight million, 64 million and one billion atoms. For single points this is acceptable: a billion-atom force evaluation of Si takes 1.8~days on one GB10, or 5.4~hours on eight with the work divided evenly (7.2~hours as run, with four GB10s running two tasks each). Every piece of every run fits on one desk-side GB10, so more units only shorten the wall time. Data-center GPUs are faster: in single-graph tests with published ALIGNN-FF models, an H200 was about 4 times faster than a GB10 with forces, and an H200 or B200 8 to 9 times faster for the energy alone, consistent with the GB10's unified memory, which the CPU and GPU share at 273~GB/s, an order of magnitude below the high-bandwidth memory of those GPUs. The GB10's advantage is not speed but that one group can own and run it continuously. For dynamics, however, the time is prohibitive: a thousand 1~fs steps of a billion atoms, one picosecond, would take about 4.9~years on one GB10. At $10^{5}$ atoms, where a force evaluation without layer checkpointing takes 3.5~s, about 25~ps of molecular dynamics per day is within reach in throughput, with a model that conserves energy, as \affs{} does and the published model does not.

In system size the billion-atom run matches the billion-atom SNAP simulation~\cite{nguyencong2021snap} with eight desk-side GPUs instead of 27,900, though for one energy and force evaluation rather than molecular dynamics. In speed it does not compete: per device, a force evaluation costs $1.6\times10^{-4}$ device-seconds per atom, about 140 times SNAP and $7\times10^{3}$ times NEP~\cite{suo2026trillion} (Table~S1). On the GB10 itself, SNAP and NEP fit 157 and 57 million Si atoms and run 11 and 780 times faster than \affs{} on one graph, while Allegro and Deep Potential stop below $10^{5}$ atoms (Figure~S1). A universal model with angular messages does far more arithmetic per atom than a potential fitted to one system, and the GB10 has 3 to 7 times less memory bandwidth than the accelerators those records used. What the GB10 offers instead is a universal model, with no system-specific training, on hardware that one group can own and run continuously.

Two limits of the implementation remain. First, for a structure stored on disk, every task holds the whole cell in host memory, 60 to 74~GB per billion atoms (62.1~GB for the polycrystal; 73.9~GB for $10^{9}$ Si atoms, measured with \aff{} before the lattice-generated pieces existed), which caps one GB10 at roughly 1.1 to 1.4 billion atoms; reading only each piece's atoms from disk would remove this, as generating pieces from the lattice already does for crystals. Second, the decomposition evaluates a fixed configuration: molecular dynamics at scale needs the pieces exchanged between steps, as in LAMMPS~\cite{thompson2022lammps}.

In summary, a universal graph neural network force field evaluates energies and forces of more than one billion atoms, two billion in the largest run, on desk-side GB10 systems, exactly wherever the 52-neighbor cap does not bind, with GPU memory set by the size of a piece rather than of the system. Polycrystals with thousands of grains and transistor-scale devices, which until now needed a supercomputer to hold, fit on hardware a single group can own. The remaining limit is time, which for now restricts the largest structures to single points. The jump in the published model's forces at its three-body cutoff, which stalls relaxation and breaks energy conservation, is removed by retraining with smooth cutoffs. Nothing in the decomposition is specific to ALIGNN-FF: it needs only per-atom energies and a finite receptive field, so other groups can run their own message-passing force fields the same way, with the halo set to their model's receptive field and the scripts released with this work. Models with long-range electrostatics, whose reach is not finite, are the exception.

\noindent\textbf{Computational Methods}\par
All calculations ran on single NVIDIA GB10 nodes, each with 128~GB of LPDDR5x memory shared by the CPU and the GPU (121.7~GiB available to PyTorch), under Slurm, with PyTorch~\cite{paszke2019pytorch} (version 2.13.0) and CUDA 13.0, the pure-PyTorch ALIGNN implementation at commit \texttt{6a20fb5} of \url{https://github.com/atomgptlab/alignn} with a patch that adds the switches described below, ASE~\cite{larsen2017ase} and matscipy neighbor lists~\cite{matscipy2023}. Production calculations are in single precision; double precision was used only for the exactness tests. \affs{} has the architecture of \aff{}: two ALIGNN and two graph-convolution layers with hidden width 128, a 5~\AA{} cutoff with at most 52 neighbors per atom and a 3.5~\AA{} three-body cutoff, which give a receptive field of 20~\AA. For the decomposition, the periodic cell of side $L$ (the code requires an orthorhombic cell) is divided into $k^{3}$ cubes of side $L/k$, with $k$ chosen so that a piece holds about a target number of atoms. Each piece contains the cube's atoms and every atom, including periodic images, within $h=20.01$~\AA{} of the cube, and is evaluated as a nonperiodic cluster in a box with more than one cutoff of vacuum. The partial energy is the sum of the per-atom energies of the owned atoms and of the bond terms whose first atom is owned. Its gradient with respect to all positions in the piece, from automatic differentiation with layer checkpointing, is added in double precision on the host to the full-cell force array. Pieces are distributed round-robin over Slurm array tasks, and a final step sums the tasks. Crystals too large to hold in host memory are generated piece by piece from the lattice, each atom's random displacement computed from a hash of its global index. Si cells have every atom displaced by a Gaussian of standard deviation 0.05~\AA. Times are the summed wall time of all tasks, from their first piece to their last, including graph construction. \affs{} multiplies every bond message by a switch that falls smoothly from 1 at 4~\AA{} to 0 at 5~\AA{}, and every angle message by the product of the switches of its two bonds (3.0 to 3.5~\AA); the published \aff{} already tapers its bond features to zero between 4 and 5~\AA, but not its angle terms at 3.5~\AA. \affs{} was trained on MatPES r2SCAN~\cite{kaplan2025matpes,furness2020r2scan} with the settings recorded for \aff{} (100 epochs, batch size 64, learning rate $10^{-3}$, energy and force weights of 1). The structures, the relaxation, the dynamics tests and further details are in the Supporting Information.

\begin{suppinfo}
Supporting Information (PDF): The Supporting Information compares this work with earlier large machine-learning force field calculations and with other machine-learning potentials on one GB10. It gives the largest single graphs of four ALIGNN-FF models, the accuracy of the decomposition against halo width, the chunked runs of all four models and the cost by structure. The extended methods include the test errors of both models, the force jumps of all published ALIGNN-FF models, the throughput against piece size and a reproduction with current software. It ends with a description of the code.

Data availability: Result files, job logs and run scripts are available at \url{https://github.com/atomgptlab/scalingmat} under \texttt{gb10/\allowbreak alignn\_\allowbreak ff\_\allowbreak max\_\allowbreak atoms/}. Structures, force arrays and per-atom energies of the largest runs are kept on the cluster; their summaries and energy maps are in the repository, with the scripts that draw the figures from them (\texttt{analysis/\allowbreak gb10\_\allowbreak billion\_\allowbreak figures.py} and \texttt{analysis/\allowbreak gb10\_\allowbreak energy\_\allowbreak maps\_\allowbreak figure.py}). The weights of \affs{} are available on Figshare~\cite{lee2026affsmooth}, and the patch that adds its switches to ALIGNN is \texttt{gb10/\allowbreak alignn\_\allowbreak ff\_\allowbreak max\_\allowbreak atoms/\allowbreak smooth\_\allowbreak messages.patch} in the repository.\end{suppinfo}

\noindent\textbf{Notes}\par
The authors declare no competing financial interest.

\begin{acknowledgement}
This work was supported by the National Science Foundation under Award No. 2607469 and the
U.S. Department of Energy under Award No. DE-SC0026725. During the preparation of this work, the authors used large language models (LLMs) to assist with drafting and editing portions of this manuscript. After using these tools, the authors reviewed and edited the content as needed and take full responsibility for the content of the published article. 
\end{acknowledgement}

\bibliography{ref}

\end{document}